\documentclass[aps,reprint,amsmath,superscriptaddress]{revtex4-1}
\usepackage{graphicx}
\usepackage{siunitx}
\usepackage{hyperref}

\begin{document}
	\title{Charge transport in oxygen-deficient EuTiO$_3$: the emerging picture of dilute metallicity in quantum-paraelectric perovskite oxides}
	\date{\today}
	\author{Johannes Engelmayer}
	\affiliation{II. Physikalisches Institut, Universit\"{a}t zu K\"{o}ln, Z\"{u}lpicher Str. 77, 50937 K\"{o}ln, Germany}
	\author{Xiao Lin}
	\altaffiliation[Now at ]{School of Science, Westlake University, 18 Shilongshan Road, 310024, Hangzhou, China}
	\affiliation{II. Physikalisches Institut, Universit\"{a}t zu K\"{o}ln, Z\"{u}lpicher Str. 77, 50937 K\"{o}ln, Germany}
	\author{Christoph P. Grams}
	\author{Raphael German}
	\author{Tobias Fr\"{o}hlich}
	\author{Joachim Hemberger}
	\affiliation{II. Physikalisches Institut, Universit\"{a}t zu K\"{o}ln, Z\"{u}lpicher Str. 77, 50937 K\"{o}ln, Germany}
	\author{Kamran Behnia}
	\affiliation{Laboratoire de Physique et d'\'{E}tude des Mat\'{e}riaux (UMR 8213 CNRS-ESPCI), PSL Research University, 10 Rue Vauquelin, 75005 Paris, France}
	\affiliation{II. Physikalisches Institut, Universit\"{a}t zu K\"{o}ln, Z\"{u}lpicher Str. 77, 50937 K\"{o}ln, Germany}
	\author{Thomas Lorenz}
	\email{tl@ph2.uni-koeln.de}
	\affiliation{II. Physikalisches Institut, Universit\"{a}t zu K\"{o}ln, Z\"{u}lpicher Str. 77, 50937 K\"{o}ln, Germany}

	\begin{abstract}
		We report on a study of charge transport in EuTiO$_{3-\delta}$ single crystals with carrier density tuned across several orders of magnitude. Comparing this system with other quasi-cubic perovskites, in particular strontium titanate, we draw a comprehensive picture of metal-insulator transition and dilute metallicity in this $AB$O$_3$ family. Because of a lower electric permittivity, the metal-insulator transition in EuTiO$_{3-\delta}$ occurs at higher carrier densities compared to SrTiO$_3$. At low temperature, a distinct $T^2$ resistivity is visible. Its prefactor $A$ smoothly decreases with increasing carrier concentration in a similar manner in three different perovskites. Our results draw a comprehensive picture of charge transport in doped quantum paraelectrics.
	\end{abstract}

	\maketitle
	 
	During the last decade, the metal-insulator transition (MIT) in weakly doped SrTiO$_3$ has attracted renewed interest. The pure compound is a highly insulating quantum paraelectric~\cite{Mueller1979}, which on the one hand becomes ferroelectric by a partial substitution of Sr by Ca (Sr$_{1-x}$Ca$_x$TiO$_3$, $\num{0.002}\leq x\leq\num{0.12}$)~\cite{Bednorz1984,DeLima2015}. On the other hand it becomes metallic upon reduction (SrTiO$_{3-\delta}$)~\cite{Spinelli2010} and even superconducting~\cite{Schooley1964} at remarkably low carrier concentrations, which identified SrTiO$_3$ as the most dilute superconductor~\cite{Lin2013}. Furthermore, a ferroelectric-like transition inside the superconducting phase has been observed in compounds with both, Ca substitution and oxygen vacancies (Sr$_{1-x}$Ca$_x$TiO$_{3-\delta}$)~\cite{Rischau2017}.
	Apart from reduction, SrTiO$_3$ has been subjected to other variants of $n$-type doping by, e.g., substituting Ti$^{4+}$ with Nb$^{5+}$ (SrTi$_{1-x}$Nb$_x$O$_3$)~\cite{Koonce1967,Binnig1980,Ohta2005,Spinelli2010}, or Sr$^{2+}$ with La$^{3+}$ (Sr$_{1-x}$La$_x$TiO$_3$)~\cite{Uematsu1984,Tang1996,Suzuki1996,Ohta2005}.
	In all three cases a $T^2$ behavior of the resistivity is found~\cite{Okuda2001,VanderMarel2011,Lin2015}.
	For many systems, the prefactor $A$ of $\rho(T)=\rho_0+AT^2$ is related to the electronic specific heat coefficient $\gamma$, since both depend on the Fermi energy $E_\mathrm{F}$, as is expressed in the Kadowaki-Woods ratio $A/\gamma^2$~\cite{Kadowaki1986}. Furthermore, $E_\mathrm{F}$ itself depends on the carrier density $n$ and one may expect a particular scaling behavior in $A(n)$ as shown for metallic SrTiO$_{3-\delta}$~\cite{Lin2015,Mikheev2016}.
	
	In order to investigate these phenomena in other systems, EuTiO$_3$ is a prime candidate, because both materials are similar in many aspects. Sr$^{2+}$ and Eu$^{2+}$ have almost the same ionic radius \footnote{In Shannon's paper of ionic radii~\cite{Shannon1976} no value is given for the ionic radius of Eu$^{2+}$ with a twelvefold coordination, but it is pointed out that the radius of Eu$^{2+}$ is only slightly larger than that of Sr$^{2+}$ for all coordination numbers. This is supported by the fact, that the lattice parameters of both compounds are almost the same~\cite{Brous1953}}. Both compounds have the ideal cubic perovskite structure (space group $Pm\bar{3}m$) at room temperature and undergo a structural phase transition to tetragonal ($I4/mcm$) upon cooling~\cite{Bussmann-Holder2011,Allieta2012}, and both are quantum paraelectrics~\cite{Mueller1979,Katsufuji2001,Kamba2007}.
	Nevertheless, there are also clear differences. SrTiO$_3$ crystals are transparent, whereas EuTiO$_3$ is black, what can be understood from band structure calculations yielding a band gap of \SI{1}{eV}~\cite{Akamatsu2011},
	whereas SrTiO$_3$ has a gap of \SI{3.2}{\electronvolt}~\cite{Baeuerle1978}. SrTiO$_3$ is non-magnetic in contrast to EuTiO$_3$ where Eu$^{2+}$ has a large, local magnetic moment of $\num{7}\mu_\mathrm{B}$. These moments order antiferromagnetically below $T_\mathrm{N}=\SI{5.5}{\kelvin}$ in a G-type configuration~\cite{McGuire1966,Scagnoli2012}.
	
	The research on $n$-doped EuTiO$_3$ is sparse. To our knowledge only five publications exist: One report deals with poly- and single-crystalline EuTi$_{1-x}$Nb$_x$O$_3$ with $x\leq\num{0.3}$~\cite{Li2015} and another two with single-crystalline Eu$_{1-x}$La$_x$TiO$_3$ ($x\leq\num{0.1}$~\cite{Katsufuji1999,Tomioka2018}).
	Studies of oxygen-deficient EuTiO$_3$ are restricted to ceramics~\cite{Kennedy2014} and thin films~\cite{Kugimiya2007}. 
	Here, we present a detailed study of single-crystalline EuTiO$_{3-\delta}$ tuned from semiconducting to metallic via reduction. 
	We derive the electron mobility and discuss its temperature dependence in comparison to that of SrTiO$_3$.
	We find an $AT^2$ resistivity behavior of metallic EuTiO$_{3-\delta}$ where $A$ systematically decreases with increasing charge-carrier content, which is discussed in a larger context of charge transport in weakly doped perovskite oxides.
	
	The EuTiO$_3$ crystals were grown by the floating-zone technique.	
	We used polycrystalline powders of Eu$_2$O$_3$ (chemical purity \SI{99.99}{\percent}), TiO (\SI{99.5}{\percent}), and TiO$_2$ (\SI{99.99}{\percent}) as starting materials. 
	The powders were mixed for \SI{1}{\hour} and the mixture was pressed to a cylindrical rod at $\SI{50}{\mega\pascal}$. In order to avoid emergence of Eu$^{3+}$ via oxygen capture, we skipped preliminary powder reactions and put the pressed rod directly into the floating-zone system.
	Centimeter-sized single crystals were grown in argon atmosphere using a growth speed of \SI{10}{\milli\meter\per\hour} and a relative rotation of the rods of \SI{30}{rpm}.
	X-ray powder diffraction measurements verified phase purity and Laue images confirmed single crystallinity.

	\begin{figure}
		\includegraphics[width=1\columnwidth]{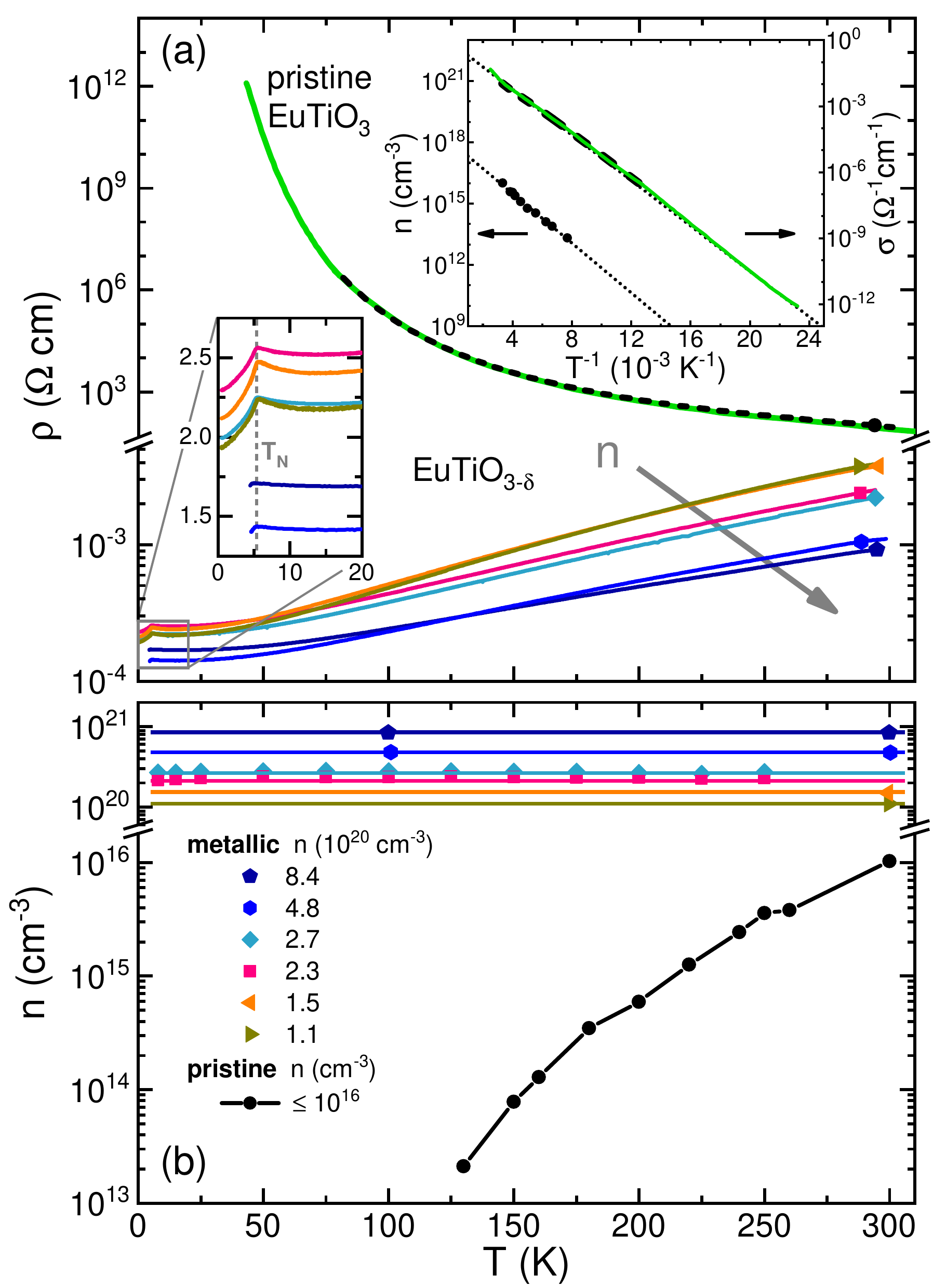}
		\caption{\label{fig:ETO-rho-vs-T} (a) Resistivity $\rho(T)$ of semiconducting pristine EuTiO$_3$ determined by DC measurements (black dashed line) and dielectric spectroscopy (green solid line) in comparison to metallic EuTiO$_{3-\delta}$ which were cut from the same single crystal and oxygen reduced. (b) Charge carrier concentrations $n$ of all samples deduced from Hall effect measurements at various temperatures. The {\color{blue} upper inset in (a) shows Arrhenius plots of $n(T)$ and of the conductivity $\sigma(T)$ together with linear fits (dotted lines). The lower inset is an enlarged view of the $\rho(T)$ anomalies at $T_\mathrm{N}=\SI{5.5}{\kelvin}$, which is $n$ independent, in agreement with a previous report~\cite{Kennedy2014}.} Note the scale breaks in both main panels.}
	\end{figure}
		
	The as-grown crystal was cut into cuboid pieces	with all faces being $\{100\}$ planes.
	In order to induce electron doping, the samples were annealed in sealed fused-quartz tubes with low argon pressure ($\lesssim\SI{E-5}{\milli\bar}$) and titanium metal powder (\SI{99.99}{\percent}) acting as oxygen catcher. The quartz tubes were heated for \SI{10}{\hour} at temperatures between \SI{650}{\celsius} and \SI{850}{\celsius} depending on the intended carrier concentration.
	In order to have an indicator for homogeneity, in each run two samples with different thicknesses (\SI{0.2}{\milli\meter} and \SI{0.4}{\milli\meter}) were annealed simultaneously in the same quartz tube.
	Resistivity and Hall effect measurements were carried out by a standard four-probe and six-probe method, respectively, using a home-built dipstick setup and a commercial $^3$He insert (Heliox, Oxford Instruments) for wet cryostats.

	Figure~\ref{fig:ETO-rho-vs-T} shows the resistivity $\rho$ and charge carrier density $n$ as a function of temperature---both in semilogarithmic scales---for different EuTiO$_{3-\delta}$ samples.
	In contrast to SrTiO$_3$, which is highly insulating, the DC conductivity of pristine EuTiO$_3$ is measurable down to about \SI{80}{\kelvin} (Fig.~\ref{fig:ETO-rho-vs-T} (a)) and is complemented with dielectric spectroscopy measurements (see Appendix A) to even lower temperature.
	Its carrier density obtained from Hall effect measurements (Fig.~\ref{fig:ETO-rho-vs-T} (b)) is temperature dependent and ranges from $n=\SI{E16}{\per\cubic\centi\meter}$ at room temperature down to $n\approx\SI{E13}{\per\cubic\centi\meter}$ at the lowest measurable temperature ($\approx\SI{130}{\kelvin}$).
	The activated behavior is clearly seen in the Arrhenius plots of both, conductivity and carrier density (upper inset of Fig.~\ref{fig:ETO-rho-vs-T}).
	The corresponding fits yield very similar activation energies (\SI{100}{\milli\electronvolt} from conductivity and \SI{120}{\milli\electronvolt} from carrier density), but both are much smaller than the theoretically expected intrinsic band gap of \SI{1}{\electronvolt}~\cite{Akamatsu2011} meaning that the pristine EuTiO$_3$ is weakly impurity-doped.
	
	To induce a MIT, the aforementioned annealing technique is used. Annealing temperatures below \SI{600}{\celsius} seem to have no effect on the oxygen content, since the $\rho(T)$ curves remain unchanged (not shown). For annealing temperatures above \SI{750}{\celsius} we obtain metallic samples with temperature-independent carrier densities that cover a range of \SI{E20}{\per\cubic\centi\meter} to \SI{E21}{\per\cubic\centi\meter} (see Fig.~\ref{fig:ETO-rho-vs-T}).
	Above \SI{130}{\kelvin} the $\rho(T)$ curves are ordered by carrier density, i.e., $\sigma$ increases upon increasing $n$ and, in reverse, the $n(T)$ curves are ordered by the high-temperature conductivity. 	
	At low temperatures, some of the $\rho(T)$ curves are crossing each other, which may partly arise from different residual resistivities and/or some uncertainty in determining the exact geometries. 
	For annealing temperatures $\SI{600}{\celsius}< T_\mathrm{ann}<\SI{750}{\celsius}$, 
	the simultaneously annealed samples of different thicknesses show large deviations in both, $\rho(T)$ and $n$.
	This indicates inhomogeneous charge carrier concentrations and thus these samples are not taken into account here. In this context, it is worth to mention that a certain gradient in the oxygen-defect concentration is naturally expected for post-annealed single crystals. However, above a certain critical concentration the wave functions of the induced charge carriers overlap sufficiently and a metallic state with an averaged homogeneous charge carrier density results.
	
	The absence of homogeneous samples between pristine and metallic EuTiO$_{3-\delta}$ hinders an exact determination of the MIT. The lowest carrier density of $\SI{E20}{\per\cubic\centi\meter}$ yields an upper boundary for the critical carrier density $n_\mathrm{c}$ of the MIT and is
	about four orders of magnitude larger than the corresponding one ($\approx\SI{E16}{\per\cubic\centi\meter}$) of SrTiO$_3$~\cite{Spinelli2010}.
	This difference can be understood by comparing the permittivities $\varepsilon$ of EuTiO$_3$ and SrTiO$_3$. While SrTiO$_3$ has an extremely large $\varepsilon$ of roughly $\num{20000}$ at low temperatures~\cite{Mueller1979}, that of EuTiO$_3$ is smaller by a factor of 50. We find $\varepsilon\approx\num{400}$ (see Appendix A) in agreement with previous single-crystal data~\cite{Katsufuji2001}, whereas smaller values are reported for ceramics~\cite{Kamba2007,Goian2009}.
	
\begin{figure}
		\includegraphics[width=1\columnwidth]{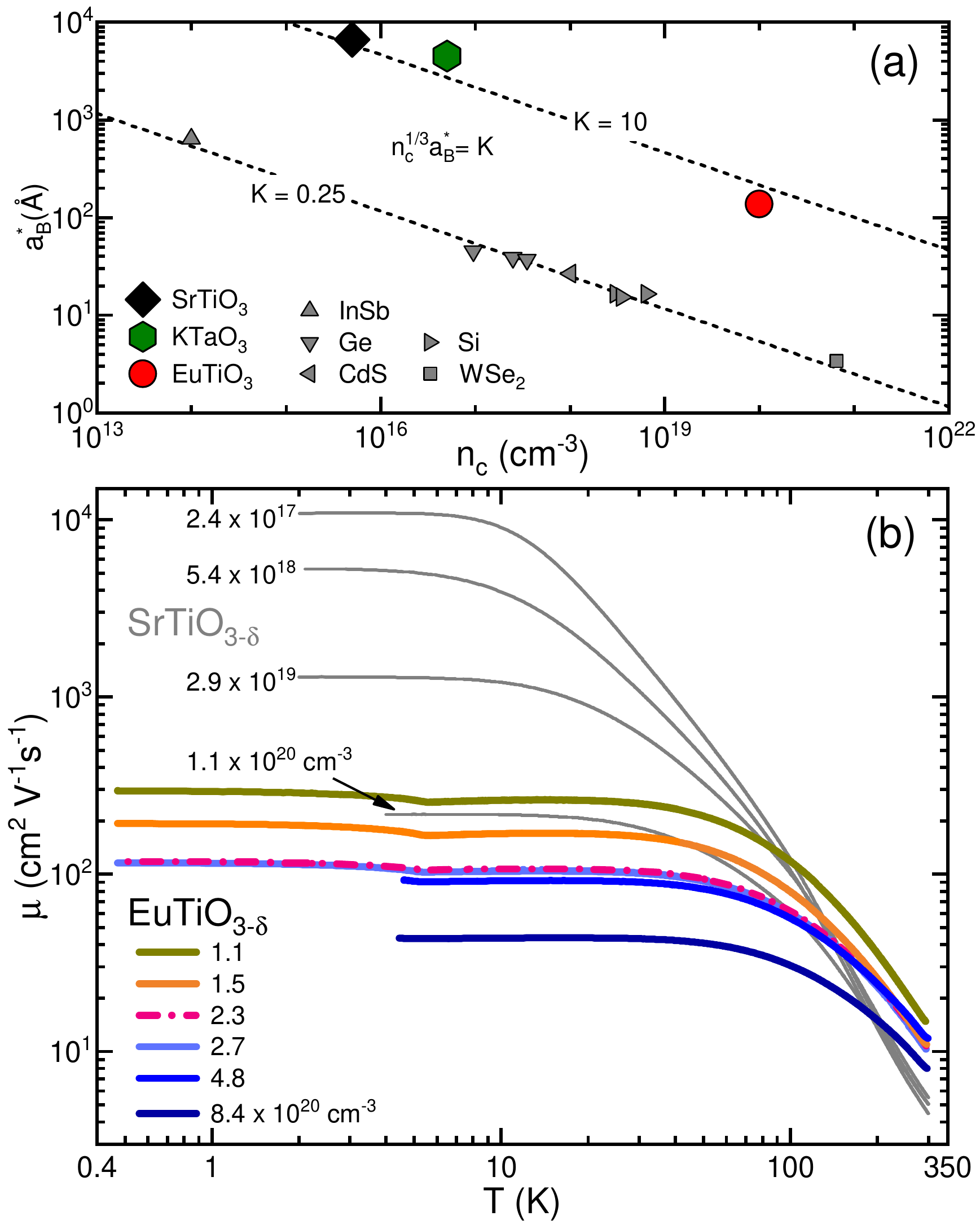}
		\caption{\label{fig:ETO-STO-mu} (a) Effective Bohr radius $a_\mathrm{B}^*$ versus critical charge carrier density $n_c$ of various doped semiconductors (taken from~\cite{Edwards1978}) in comparison to the observed MIT of EuTiO$_{3-\delta}$ and related doped oxides SrTiO$_3$~\cite{Spinelli2010} and KTaO$_3$~\cite{Wemple1965,Uwe1979}. Dashed lines represent the scaling behavior $n_c^{1/3}a_\mathrm{B}^*=K$ with different values of $K$. {\color{blue} (b) Mobility $\mu(T,n) $ of the metallic EuTiO$_{3-\delta}$ samples in comparison to that of SrTiO$_{3-\delta}$.}}
	\end{figure}

	Of course, these values were obtained for pristine EuTiO$_3$. For doped samples, one defines an effective Bohr radius $a_\mathrm{B}^*=a_\mathrm{B}\varepsilon m_e/m^*$, which renormalizes $a_\mathrm{B}\approx\SI{0.5}{\angstrom}$ of the hydrogen atom by taking into account the permittivity $\varepsilon$ and the band mass $m^*$.
	The so-called Mott criterion~\cite{Mott1961} compares $a_\mathrm{B}^*$ as a measure for the overlap of the electronic wave functions to the average distance between donor atoms $n^{-1/3}$.
	The huge low-temperature $\varepsilon$ of SrTiO$_3$ results in an effective Bohr radius of about \SI{6700}{\angstrom}, compared to $a_\mathrm{B}^*\approx\SI{130}{\angstrom}$ for EuTiO$_3$. Here, we use $m^*=\num{1.5}m_e$ as determined for the lowest lying conduction band of SrTiO$_{3-\delta}$~\cite{Lin2014} for both SrTiO$_3$ and EuTiO$_3$. The much smaller value of $a_\mathrm{B}^*$ explains that $n_c$ of EuTiO$_3$ is about four orders of magnitude larger than that of SrTiO$_3$. In passing, we also note that the influence of the above-mentioned inhomogeneities in the oxygen-defect concentrations is suppressed more rapidly with increasing $a_\mathrm{B}^*$.
	Figure~\ref{fig:ETO-STO-mu} (a) shows the scaling behavior $n_c^{1/3}a_\mathrm{B}^*=K$ as dashed lines for different values of $K$.
	Experiments on doped semiconductors have detected a sharp MIT at a critical density of $n_c$ and the available data follows a scaling relation with $K=\num{0.25}$~\cite{Edwards1978,Edwards1995}, which corresponds to the so-called Mott criterion~\cite{Mott1961}. In perovskite oxides, there is no experimental data resolving a sharp MIT at $n_c$ and metallicity is observed in EuTiO$_3$, SrTiO$_3$~\cite{Spinelli2010}, and KTaO$_3$~\cite{Wemple1965,Uwe1979} at carrier densities which are much larger than expected according to the Mott criterion. Nevertheless, these carrier densities scale with $a_\mathrm{B}^*$.

Figure~\ref{fig:ETO-STO-mu}~(b) displays the mobility $\mu=1/(ne\rho)$ of metallic EuTiO$_{3-\delta}$ as a function of temperature in double-logarithmic scales. Below $\SI{40}{\kelvin}$ all $\mu(T)$ curves approach constant values, which are ordered by carrier density $n$, i.e., $\mu(n)$ systematically decreases with increasing $n$.
The additional kinks result from the magnetic order at $T_\mathrm{N}=\SI{5.5}{\kelvin}$ as already shown in Fig.~\ref{fig:ETO-rho-vs-T}~(a) for $\rho(T)$.
In the high-temperature regime, the mobility curves decrease due to increasing electron-phonon scattering and seem to approach an $n$-independent power law. Such a behavior has been already observed in SrTiO$_3$~\cite{Lin2017}. For comparison, we also show the mobility data 
of four SrTiO$_{3-\delta}$ crystals with $\SI{E17}{\per\cubic\centi\meter}\leq n\lesssim\SI{E20}{\per\cubic\centi\meter}$.
Because SrTiO$_{3-\delta}$ is already metallic for very low carrier densities, higher mobilities than in EuTiO$_{3-\delta}$ are reached in the low-temperature regime, but even across both compounds all curves remain ordered by increasing $n$.
Towards high temperature, the mobility curves $\mu(T,n)$ of SrTiO$_{3-\delta}$ merge and fall below those of EuTiO$_3$ above about \SI{200}{\kelvin}. This is surprising in view of the structural phase transition of EuTiO$_{3}$, which is in that temperature range~\cite{Kennedy2014, Bussmann-Holder2011}. In contrast, the transition in SrTiO$_{3-\delta}$ appears at $T_s\simeq\SI{105}{\kelvin}$~\cite{Mueller1968,Mueller1979} and linearly decreases with increasing charge-carrier content~\cite{Tao2016}. Using X-ray and Raman scattering measurements, we derive $T_s\simeq\SI{260}{\kelvin}$ on our pristine EuTiO$_3$ and $T_s\simeq\SI{200}{\kelvin}$ for the highest $n=\SI{8.4E20}{\per\cubic\centi\meter}$ (to be published elsewhere). 
However, neither SrTiO$_{3-\delta}$ nor EuTiO$_{3-\delta}$ show any anomalies in the mobility data reflecting the structural transitions. 
Recently, both the magnitude and temperature dependence of the mobility in SrTiO$_{3-\delta}$ have attracted attention~\cite{Mishchenko2018,Zhou2018}. Mischenko et al.~\cite{Mishchenko2018} argue that a polaronic approach can lead to a scattering rate larger than the thermal energy of carriers in agreement with the data. Ab initio calculations by Zhou et al.~\cite{Zhou2018} reproduce the experimentally observed $T^{-3}$ temperature dependence of the mobility of SrTiO$_{3-\delta}$~\cite{Lin2017}, but the calculated absolute value is an order of magnitude larger than the experimental data. Moreover, in these theoretical approaches the antiferrodistortive soft mode does not play a key role, in agreement with the absence of anomalies in the measured mobility data.

	\begin{figure}
		\includegraphics[width=1\columnwidth]{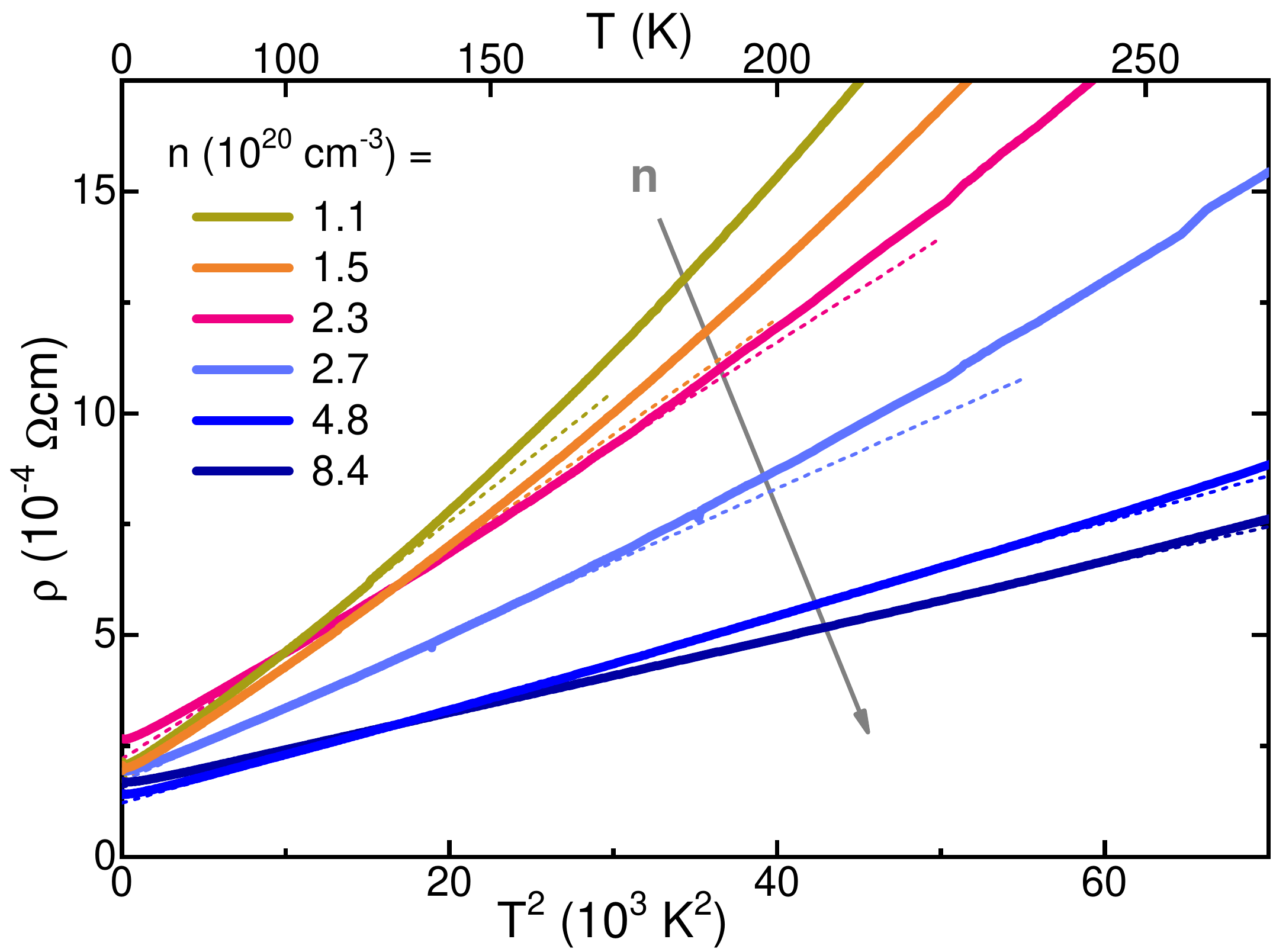}
		\caption{\label{fig:ETO-rho-vs-T2} Resistivity $\rho$ of EuTiO$_{3-\delta}$ as a function of $T^2$. The dashed lines are fits of the form $\rho(T)=\rho_0+AT^2$. With increasing $n$ the prefactor $A$ decreases and the temperature range of the $T^2$ behavior increases.}
	\end{figure}	

\begin{figure}
		\includegraphics[width=1\columnwidth]{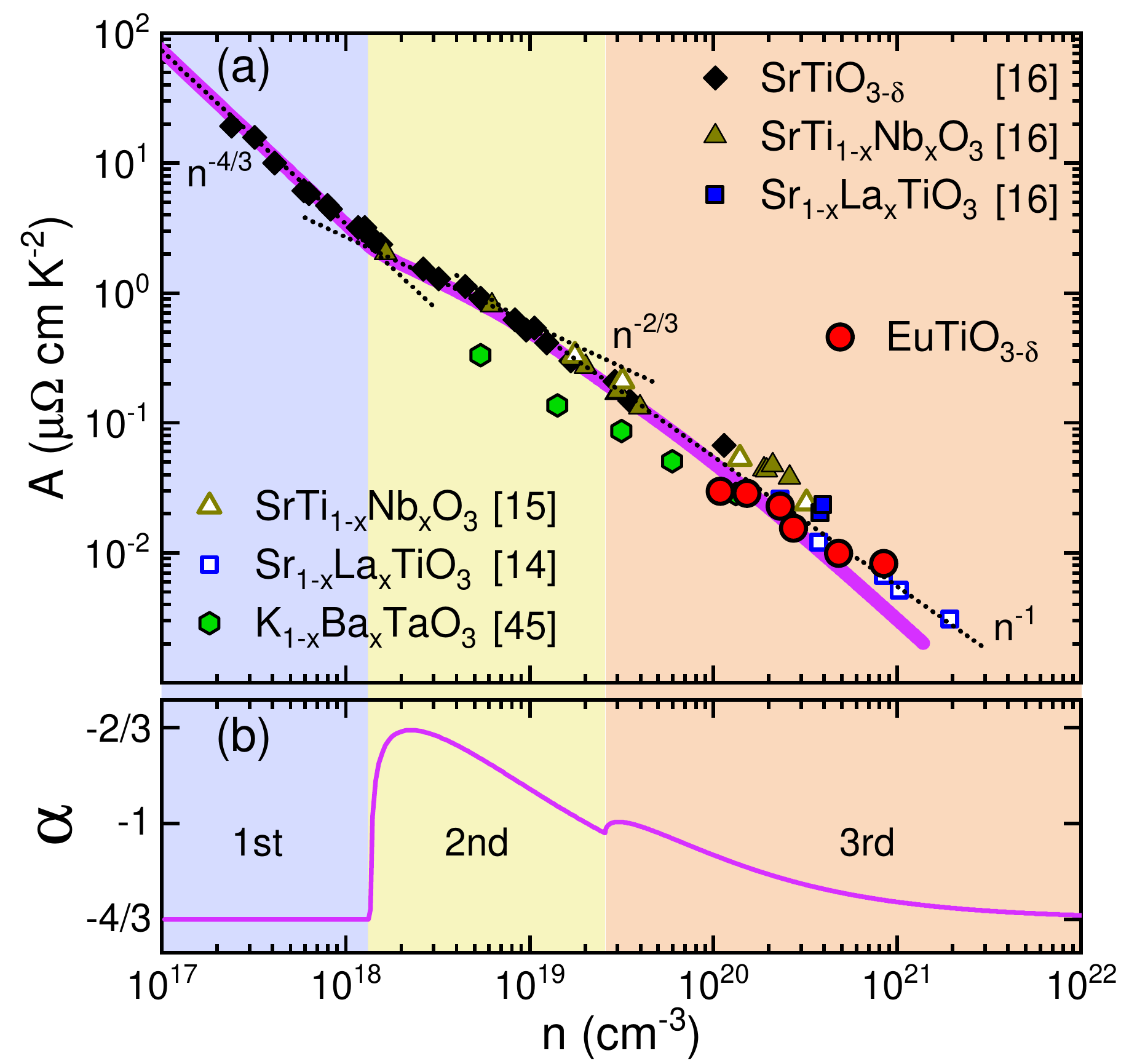}
		\caption{\label{fig:ETO-A-vs-n} (a) Prefactor of the $AT^2$ resistivity versus carrier concentration $n$ in doped perovskites EuTiO$_3$, SrTiO$_3$~\cite{Lin2015,VanderMarel2011,Okuda2001}, and KTaO$_3$~\cite{Sakai2009}. Dotted black lines are guides to the eye. The thick line in (a) represents $A(n)$ calculated for a three-band model, the corresponding exponent $\alpha$ of $A\propto n^\alpha$ is shown in (b) (see text for details). Color boundaries indicate band edges of doped SrTiO$_3$~\cite{Lin2014}.}
	\end{figure}

	Figure~\ref{fig:ETO-rho-vs-T2} shows the resistivity $\rho$ as a function of $T^2$ together with fits of the form $\rho(T)=\rho_0+AT^2$ (dashed lines).
	The fits deviate from the data for high temperatures and with increasing carrier density the temperature range of the $T^2$ behavior systematically increases, which is in agreement with the findings for SrTiO$_{3-\delta}$~\cite{Lin2015}.
	In EuTiO$_{3-\delta}$ we have an additional deviation at low temperatures that is related to the magnetic transition at $T_\mathrm{N}=\SI{5.5}{\kelvin}$.
	Figure~\ref{fig:ETO-A-vs-n} (a) shows the prefactor $A$ from these fits as a function of $n$ in double-logarithmic scales.
	Here, we compare $A(n)$ for EuTiO$_{3-\delta}$ to that of SrTiO$_{3-\delta}$, Sr$_{1-x}$La$_x$TiO$_3$, and SrTi$_{1-x}$Nb$_x$O$_3$~\cite{Lin2015,Okuda2001,VanderMarel2011}, and we also include $A(n)$ of the non-titanate perovskite K$_{1-x}$Ba$_x$TaO$_3$~\cite{Sakai2009}. 
	All titanate systems follow a general trend as is marked by dotted black lines, which are guides to the eye and indicate power laws $A\propto n^\alpha$ with $\alpha=-4/3,-2/3,-1$.
	Band structure calculations for $n$-doped SrTiO$_3$~\cite{VanderMarel2011} yield a model with three bands that are filled consecutively with increasing $n$. 
	The critical carrier densities $n_{c1}$ ($n_{c2}$), at which the filling of the second (third) band sets in, are known from experiments~\cite{Lin2014} and illustrated by background-color boundaries.
	Below $n_{c1}$, where only the first band is filled, a power law $n^{-4/3}$ is seen as is expected for a single parabolic band with $E_\mathrm{F}\propto n^{2/3}$ and a simple $A\propto E_\mathrm{F}^{-2}$ relation.
	When the second band starts to be filled at $n_{c1}$, the exponent $\alpha$ of $A\propto n^\alpha$ suddenly increases and finally approaches $-1$, which does not change much above $n_{c2}$.

	The increase of $\alpha$ is a natural consequence of a three-band system. If we consider the most simple case of three parabolic bands with band minima at energies $E_i$, effective masses $m_i$, and densities of states $g_i(E)\propto m_i^{3/2}\sqrt{E-E_i}$, then each band contributes
	\begin{equation}
		n_i(E_\mathrm{F})=\frac{1}{3\pi^2}\left(\frac{2m_i}{\hbar^2}\right)^{3/2}\int_{E_i}^{E_\mathrm{F}}\sqrt{E-E_i}\,\mathrm{d}E
	\end{equation}
	to the total electron density $n(E_\mathrm{F})=\sum_in_i(E_\mathrm{F})$.
	We use the band masses $m_0=m_2=\num{1.5}m_e$ and $m_1=\num{3.5}m_e$ from Shubnikov-de-Haas measurements of SrTiO$_{3-\delta}$~\cite{Lin2014} and adjust $E_{1,2}$ to \SI{2}{\milli\electronvolt} and \SI{10}{\milli\electronvolt}, respectively, to match the experimental critical carrier densities \footnote{In the three-band model of van der Marel et al.~\cite{VanderMarel2011} the lowest band is heavy, while the others are light. However, in our much simpler model with purely parabolic bands, we neglect band repulsion. Thus, the second band is the heavy one and crosses the lighter first band.}.
	From the inverse function $E_\mathrm{F}(n)$ we calculate $A(n)\propto E_\mathrm{F}^{-2}(n)$ which
	describes the data of the doped titanates over almost the entire range of $n$, as is shown by the thick line in Fig.~\ref{fig:ETO-A-vs-n} (a).
	This also holds for the exponent $\alpha$ of $A\propto n^\alpha$ obtained from the slope of $\log A$ vs. $\log n$ (Fig.~\ref{fig:ETO-A-vs-n} (b)).	
	In view of the simple model, which neglects deviations from the parabolic band shapes as well as their anisotropy, this good agreement with the experimental data is remarkable.
	The available $A(n)$ data of the non-titanate perovskite K$_{1-x}$Ba$_x$TaO$_3$~\cite{Sakai2009} fit into this picture as well, because this material has lower effective masses ($\num{0.55}m_e$ to $\num{0.8}m_e$)~\cite{Uwe1979}. Consequently, at a given carrier concentration $n$, the Fermi energy is larger and the prefactor $A$ is lower compared to the titanates.
	A more sophisticated theoretical treatment could provide a generalized uniform description of the $A(n)$ behavior for an even larger variety of metallic perovskite oxides with low carrier densities.

	In summary, we present a detailed report of the metal-insulator transition in oxygen-deficient single-crystalline EuTiO$_3$, which shows many similarities with that in SrTiO$_3$. However, it sets in at a much higher carrier concentration (factor \num{E4}), which results from the smaller permittivity of EuTiO$_3$, implying a smaller effective Bohr radius $a_\mathrm{B}^*$, i.e., a smaller overlap of the electronic wave functions.
	We show that metallicity in three perovskite oxides scales with the effective Bohr radius $a_\mathrm{B}^*$, but it emerges at a carrier density much larger than suggested by the Mott criterion.
	The low-temperature mobility of metallic EuTiO$_3$ and SrTiO$_3$ systematically increases with decreasing charge carrier concentration across both materials.
	We find an $AT^2$ behavior in $\rho(T)$ of metallic EuTiO$_{3-\delta}$ where the prefactor $A(n)$ systematically decreases with increasing charge carrier density $n$ and even quantitatively agrees with $A(n)$ of doped SrTiO$_3$. This general behavior of $A(n)$ can be described within a three-band model.

\appendix

	\section{Dielectric Spectroscopy}

The dielectric and transport properties of pristine EuTiO$_3$ towards higher resistivities were determined by contact based impedance spectroscopy. These measurements were performed in a commercial $^4$He-flow cryo-magnet ({\sc Quantum-Design PPMS}) on crystals in capacitor geometry with metallized surfaces $A\approx\SI{4}{\milli\meter\squared}$ and thickness $d\approx\SI{0.5}{\milli\meter}$ along a cubic [100] axis.
We used a high-impedance frequency response analyzer ({\sc Novocontrol}) and a vector network analyzer ({\sc ZNB8, Rohde\&Schwarz}) to cover a joint frequency range $\SI{1}{\hertz}\leq\nu\leq\SI{100}{\mega\hertz}$ with voltage stimulation below 1~V$_\mathrm{rms}$.

\begin{figure}
	\includegraphics[width=\columnwidth]{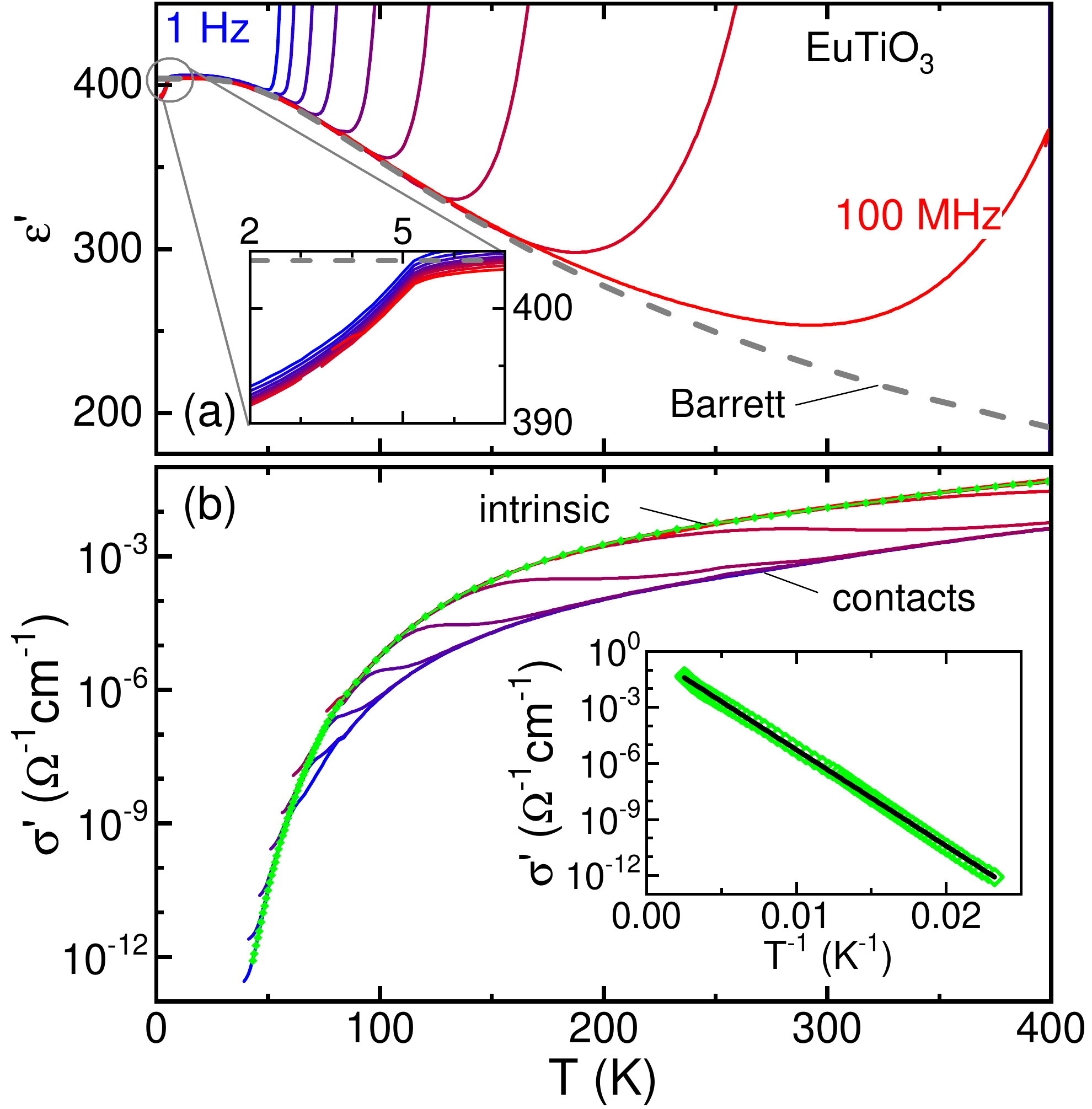}
	\caption{\label{fig:ETO-sigma-eps}(a) Temperature dependent $\varepsilon'$ measured for frequencies $\SI{1}{\hertz}\leq\nu\leq\SI{100}{\mega\hertz}$. The steep rise of $\varepsilon'(T,\nu)$ for high-$T$/low-$\nu$ results from contact contributions. The intrinsic low-$T$/high-$\nu$ behavior of $\varepsilon'(T,\nu)$ is fitted via the Barrett formula (dashed line)~\cite{Barrett1952}. The inset shows the anomaly of $\varepsilon'(T,\nu)$ at $T_\mathrm{N}=\SI{5.5}{\kelvin}$. (b) Corresponding conductivity data $\sigma'(T)$. The inset shows an Arrhenius-plot of the intrinsic $\sigma'(T)$ with a linear fit (black line).}
\end{figure}

As expected for semi-conductors, Schottky-type depletion layers at the contact interfaces cause a capacitive contribution $C_C$, which together with the contact resistance $R_C$ form 
an $RC$ element in series with the intrinsic sample impedance. This gives rise to Maxwell-Wagner-type relaxational effects~\cite{Maxwell1954}, but for frequencies $2\pi\nu> 1/R_CC_C$ the contacts effectively are short-circuited~\cite{Lunkenheimer2002,Niermann2012}. 
The crossover from contact-dominated to intrinsic response is clearly seen in the permittivity $\varepsilon$ as well as in the conductivity $\sigma$. The low-$T$/high-$\nu$ limit of the frequency- and temperature-dependent data represents the intrinsic quasi-static $\varepsilon$ (Fig.~\ref{fig:ETO-sigma-eps} (a)). The corresponding intrinsic $\sigma$, marked in green in Fig.~\ref{fig:ETO-sigma-eps}~(b) agrees well the inverse DC resistivity $1/\rho_\mathrm{DC}$ (see Fig.~\ref{fig:ETO-rho-vs-T}~(a)).

EuTiO$_3$ is a quantum paraelectric where long-range order is prevented by quantum fluctuations. The fingerprint of quantum-paraelectric behavior is a Curie-like rise of the permittivity with decreasing temperature followed by a saturation at an elevated $\varepsilon(T\rightarrow 0)$, which can be modeled by the well known Barrett formula~\cite{Barrett1952}
\begin{equation}
\varepsilon(T)= \frac{C}{(T_\Omega/2)\coth(T_\Omega/2T)-T_0} + \varepsilon_\infty\,.
\end{equation}
Here, $T_\Omega$ represents the influence of quantum fluctuations and $T_0$ is the paraelectric Curie-temperature. The fit of the high-frequency data of $\varepsilon(T<\SI{200}{\kelvin})$ reveals $T_\Omega\simeq\SI{160}{\kelvin}$ and $T_0\simeq \SI{-190}{\kelvin}$. The value of $T_\Omega$ agrees with a previous report~\cite{Katsufuji2001} and, remarkably, it is four times larger compared to SrTiO$_3$~\cite{Mueller1979,Hemberger1996} indicating much stronger quantum fluctuations in EuTiO$_3$. Our $T_0$ value differs in magnitude from~\cite{Katsufuji2001} where a considerably smaller temperature range could be evaluated, but is also negative denoting rather antiferroelectric correlations in EuTiO$_3$. As shown in the inset of Fig.~\ref{fig:ETO-sigma-eps}~(a), $\varepsilon(T)$ has a clear anomaly at $T_\mathrm{N}=\SI{5.5}{\kelvin}$, which results from a significant magneto-electric coupling~\cite{Katsufuji2001}.

\begin{acknowledgments}
	We acknowledge support by the DFG (German Research Foundation) via project number 277146847 - CRC
	1238 (Subprojects A02, B01, B02 and B03). This work is part of a DFG-ANR project funded by Agence Nationale de la Recherche \mbox{(ANR-18-CE92-0020-01)} and by the DFG through projects {LO~818/6-1} and \mbox{HE~3219/6-1}. X.~L. acknowledges support by the Alexander von Humboldt Foundation and Zhejiang Provincial Natural Science Foundation of China under Grant No. LQ19A040005.
\end{acknowledgments}


%\bibliography{EuTiO3}
	
\end{document}